TITLE: Vital Signs: Seismology of icy ocean worlds


AUTHORS:

Steven D. Vance[1*], Sharon Kedar[1], Mark P. Panning[2], Simon C. Stähler[3,4], Bruce G. Bills[1], Ralph D. Lorenz[5], Hsin-Hua Huang (黃信樺)[6,7], W. T. Pike[8], Julie C. Castillo, Philippe Lognonné[9], Victor C. Tsai[7], Alyssa R. Rhoden[10]

AFFILIATIONS:

[1] Jet Propulsion Laboratory California Institute of Technology, Pasadena, California.

[2] Department of Geological Sciences, University of Florida, Gainesville, Florida.

[3] Dept. of Earth and Environmental Sciences, Ludwig-Maximilians-Universität (LMU), Munich.

[4] Leibniz Institute for Baltic Sea Research (IOW), Rostock.

[5] Johns Hopkins Applied Physics Laboratory, Laurel, Maryland.

[6] Institute of Earth Sciences, Academia Sinica, Taipei.

[7] Seismological Laboratory, California Institute of Technology, Pasadena, California.

[8] Optical and Semiconductor Devices Group, Department of Electrical and Electronic Engineering, Imperial College, London

[9] Univ Paris Diderot-Sorbonne Paris Cité, Institut de Physique du Globe de Paris, Paris. Engineering, Imperial College, London.

[10] School of Earth and Space Exploration, Arizona State University, Tempe, Arizona.

*Corresponding Author: svance@jpl.nasa.gov, MS 321-560, 4800 Oak Grove Drive, Pasadena, California, 91109, USA.





ABSTRACT:

Ice-covered ocean worlds possess diverse energy sources and associated mechanisms that are capable of driving significant seismic activity, but to date no measurements of their seismic activity have been obtained. Such investigations could probe their transport properties and radial structures, with possibilities for locating and characterizing trapped liquids that may host life and yielding critical constraints on redox fluxes, and thus on habitability. Modeling efforts have examined seismic sources from tectonic fracturing and impacts. Here, we describe other possible seismic sources, their associations with science questions constraining habitability, and the feasibility of implementing such investigations. We argue, by analogy with the Moon, that detectable seismic activity on tidally flexed ocean worlds should occur frequently.  Their ices fracture more easily than rocks, and dissipate more tidal energy than the <1 GW of the Moon and Mars. Icy ocean worlds also should create less thermal noise for a due to their greater distance and consequently smaller diurnal temperature variations. They also lack substantial atmospheres (except in the case of Titan) that would create additional noise. Thus, seismic experiments could be less complex and less susceptible to noise than prior or planned planetary seismology investigations of the Moon or Mars.


# 1. Introduction

*The ice was here, the ice was there*,

*The ice was all around*:

*It cracked and growled, and roared and howled,*

*Like noises in a swound!*

       -The Rime of the Ancient Mariner (1797) by Samuel Taylor Coleridge

The coming years and decades could see the development and launch of a series of missions to explore the ocean worlds of the solar system. Space missions—*Voyager, Galileo, Cassini, New Horizons* (Kohlhase, 1977; Russell, 2012; Stern, 2009)—and ground based observations have returned strong evidence for salty oceans within Europa, Ganymede, Callisto, Enceladus, and Titan, and indications of potential oceans or partially molten regions in Triton and Pluto (Nimmo and Pappalardo, 2016), Dione (Beuthe et al., 2016) and Ceres (Ruesch et al., 2016).  The primary goals of ocean world missions would be to characterize the habitability of the most promising bodies and to search for life. Their interiors hold the clues for determining their thermal and chemical make-up and thus their habitability.

Planetary interiors have been investigated mainly with combined gravity and magnetic field mapping. However, these techniques cannot unambiguously retrieve structural boundaries (petrological or mechanical). Radar sounders can provide information on such transitions, but their signals can penetrate to only tens of km at most, due to scattering and dielectric absorption. The technique that can most efficiently reveal the detailed structures of planetary interiors is seismology.  Ultra-sensitive seismometers are critical for detecting faint motions deep within the planet and activity closer to the surface. Such motions can be used to



determine interior density structure while also revealing active features such as plate tectonics, volcanism, oceanic and ice flow, and geyser-like eruptions. These broad applications of planetary seismology have been well explored at solid silicate bodies, including the Moon, Mars, and Venus (e.g., Lognonné, 2005; Knapmeyer, 2009). In this paper, we focus on assessing the habitability of ocean worlds by listening for distinct "vital signs" of present day activity: fluid motion in the shallow subsurface, seismic signals emanating from cryovolcanos, and internal ocean circulation, by analogy with recent developments in cryoseismology on Earth (Podolskiy and Walter 2016).

Seismology could aid in understanding the deposition of materials on the icy surfaces of ocean worlds and their exchange with the underlying oceans. Extrusion of brines onto the surface (Collins and Nimmo, 2009; Kattenhorn and Prockter, 2014) may create conditions analogous to geyser formation and eruption on Earth, which have in recent years been shown to have distinct seismic characteristics (e.g., Kedar et al., 1996; 1998). By constraining the rate of fracturing and fluid motion within the surface ice, seismology could establish better bounds on the rate of overturn of the surface, which on Europa has been linked to the flux of oxidized materials into the underlying ocean (Hand et al., 2007, 2009; Greenberg, 2010; Pasek and Greenberg, 2012; Vance et al., 2016). Moreover, the distinct seismic profile of the ocean constrains the salinity and pH of the ocean, and thus the redox flux integrated through time (Vance et al. 2017). Seismic measurements can help to reveal the extent to which surface hydrocarbons on Titan interact with the near subsurface (Hayes et al., 2008) and underlying aqueous ocean (e.g., Fortes, 2000; Fortes et al., 2007; Grindrod et al., 2008). Europa's layered linear fractures and bands are possibly organized by plate tectonics and subduction (Greenberg



et al., 1998; Kattenhorn and Prockter, 2014), suggesting different mechanisms of fracturing that may also occur on other worlds. Each of these should have unique seismic signatures. Porosity gradients (Nimmo and Manga, 2009; Aglyamov et al. 2017) may be exploited—via their distinct wave speeds, attenuation, and anisotropy—as a measure of near-surface fracturing, but are also sources of scattering and seismic attenuation of weaker signals from the deeper interior. While planned mapping and radar may establish the distribution of fluids and the connection of fractures to the deeper interior, only seismology can identify deeper interfaces between fluids and solids.

The large satellites, Ganymede, Callisto, and Titan, contain oceans extending hundreds of km into their interiors (Vance et al. 2014; 2017). These deep ocean worlds are intriguing targets for astrobiology because of the possibility for remnant heat and internal activity, and also because the high pressures in their interiors may provide clues to the nature of volatile-rich exoplanets. Dense high-pressure ices (III, V, VI) can cover the lower portions of their oceans (e.g., Poirier, 1982). Experimental measurements of phase equilibria (Hogenboom et al., 1995), thermodynamics in salty fluids (Vance and Brown, 2013), and thermal models (Vance et al., 2014) support the idea that brine layers can occur underneath and between high-pressure ice layers. For such layered oceans to occur would require increased internal heating in the rocky interior after the formation of high-pressure ices. The related phenomenon of "upward snow"—buoyant high-pressure ices in the lower part of a salty ocean—requires very high salinity that likely occurs only as the ocean nears complete freezing, as may be the case for Callisto (Vance et al., 2017). Fluids moving within the high-pressure ices may govern heat transport through multi-phase convection (Choblet et al. 2017; Kalousova et al. *personal*



*communication*) and porous flow (Goodman, 2016).  Seismology is the only practical means for determining the thicknesses of these high-pressure ice layers, their temperature structure and thus their geodynamic state, and the possible presence of fluids within and between them.

Here, we offer a broad assessment of previously unconsidered seismic sources on icy ocean worlds. We do not quantitatively model signal propagation, but instead point to work published elsewhere (Panning et al., 2017; Stähler et al., 2017), including detailed assessments of the possible interior structures and seismic properties (Vance et al, 2017).  For currently known ocean worlds, we consider science objectives for future seismic investigations, possible implementations, and technical challenges.  We suggest priorities based on feasibility and potential science return. To motivate the astrobiological applications of planetary seismology, we first describe seismic characteristics of known and candidate ocean worlds.

## 2. Seismic Characteristics of Icy Ocean Worlds

Figure 1 summarizes likely seismic sources for Europa, and their estimated acceleration spectral density (in acceleration per root Hz) versus frequency based on prior published investigations. The response characteristics of currently available instrumentation are also shown.

### 3.3. Tides as a key source of seismic activity

Frequent and energetic seismic activity should occur on all known ocean worlds, based on the estimated tidal dissipation. Table I shows the comparative tidal dissipation estimated for different targets. Tidal dissipation in the ice exceeds that in the Moon for all known ocean worlds (Table I).  If the ice is thicker than a few km, tidal dissipation is probably larger in the ice than in the rocky interior; of Europa's estimated dissipation of ~1000 GW (Tobie et al., 2003;



Hussmann and Spohn, 2004; Vance et al., 2007) only 1-10 GW is expected in the rocky mantle. While modest in comparison to dissipation in the ice, this exceeds the Moon's 1 GW of tidal energy (the nominal minimum for Europa's mantle). Tidal deformation of the Moon generates continuous ground motion with intensity following the tidal period (Goins et al., 1981). For homogeneous small bodies such as the Moon or the rocky interiors of ocean worlds, the predicted lag between tidal and seismic dissipation is negligible (Efroimsky, 2012), so it should be expected that deformation on icy ocean worlds will also create ground motions correlated with the tides.

## Table I here

### 3.4. Seismic sources within the ice

#### 2.2.1. Impulsive events

Several prior seismic studies have considered the seismic signals due to large surface fractures on Europa (Kovach and Chyba, 2001; Lee et al., 2003; Cammarano et al., 2006; Panning et al., 2006). These studies focused on broadband signals in the 0.001 to 10 Hz range and trapped waves within the ice. The formation of fractures at Europa and to a lesser extent Ganymede and Enceladus has been modeled in detail (Lee et al., 2003, 2005; Kattenhorn and Hurford, 2009; Nimmo and Manga, 2009; Rhoden et al., 2012; Walker and Schmidt, 2015). The energy release and frequency of occurrence on those bodies are unknown, but should be expected to generate similar waveforms on the different worlds (Stähler et al. 2007). *Cassini* tracking data have exposed a time-variable component to Titan's gravity field (Iess et al., 2012) indicating that "Titan is highly deformable over time scales of days", with Love number $k_2$~0.6 indicating the presence of a deep ocean. As at Enceladus, this deformation is likely to manifest



in the generation of seismic signals, for example generated at the base of the crust (Mitri and Showman, 2008). On Pluto, seismic sources may be limited due to a lack of tidal deformation to drive activity. Hypothesized convection of volatile ices within Sputnik Planitia (McKinnon et al. 2016) may cause detectable seismicity, but the ice layer may be too thin for a seismometer to probe the detailed structure at the base of the deposit. Similarly, on Triton, fracturing due to ongoing tidal deformation and convection within the surface ice layer, and the cryovolcanic eruptions of nitrogen observed by *Voyager* 2 (Soderblom et al., 1990) could produce seismic energy, enabling characterization of Triton's interior structure and detection of an ocean layer, should it exist. Chaotic terrains, sills, pits, and domes observed on Europa are also candidate sources (Collins and Nimmo, 2009; Michaut and Manga, 2014; Culha and Manga, 2016). Their frequencies of occurrence and corresponding energy release are uncertain.

Impacts can also cause observable ground motions. They are probably too infrequent on Europa to be a source of seismic events on the notional ~month-long timescale of a lander (0.002–5 predicted direct P detections per year; Tsuji and Teanby, 2016). For a short-duration mission, impacts are probably no better a source of seismic information on other ocean worlds. The impact probability for Ganymede is roughly double that for Europa, but the expected impact speeds are ~30% smaller (Zahnle et al., 2003), reducing the kinetic energy (as $v^2$) and therefore the expected seismic energy. Callisto, more distant from Jupiter, has a similar impact probability as Europa, but with even smaller impact speeds (40% less). The impact rate and impact speeds on Titan are similar or less than those on Callisto, and in any case impactors would be a less effective seismic source owing to atmospheric breakup and wind noise. Triton has a similar rate of impacts as Europa (Zahnle et al., 2003).



The propagation of waves due to impulsive seismic events, which has been simulated in detail for Europa, Ganymede, Enceladus, and Titan (Stähler et al., 2017), is dictated by velocity variations with depth, and the strong velocity contrast between ice and ocean. These will create trapped shear-horizontal (SH; Love) and shear-vertical (SV; Rayleigh) surface waves. At relatively high frequencies, Rayleigh waves do not sense the bottom of the ice shell, and the small depth dependence of velocity within the ice shell leads to a non-dispersive Rayleigh wave with equal phase and group velocity. At longer periods, the waves interact with the bottom of the ice shell and transition into a flexural mode with phase velocity proportional to the square root of frequency and group velocity higher than phase velocity by a factor of 2. The transition between these two types of surface waves produces a group velocity peak at a characteristic frequency between 0.1 and 0.01 Hz that depends primarily on the ice shell thickness (Panning et al., 2006). Another SV mode (Crary, 1954), propagates by constructive interference of internal reflections within the ice. This constructive interference occurs only at a characteristic frequency, $f$, which is a function of the ice shell thickness $H$ and wave velocities $V_S$ and $V_P$ in the ice, $f=V_S/(H \cos \theta_{cr})$, where the critical angle is defined as $\theta_{cr}=sin^{-1}(V_S/V_P)$. For a 20 km ice shell, f≈0.11 Hz, and for a 5 km ice shell f≈0.44 Hz. The wave propagates at a horizontal phase velocity of $V_S/sin \theta_{cr} \approx V_P$, and therefore arrives after the first arriving compressional waves from the ice, but for distances of less than 30°, before mantle phases and much precedes the direct SH waves. A 3-axis seismometer monitoring Rayleigh, Love and Crary waves from 0.01-10 Hz might determine ice thicknesses in the range 5-50 km, localize events, and characterize velocity dispersion (Kovach and Chyba, 2001; Lee et al., 2003; Panning et al., 2006; Stähler et al., 2017).



# Figure 1 here

### 2.2.2. Free Oscillations

The measurement of very long period surface displacements has been proposed for detecting the quasi-elastic deformation of Europa's crust (Korablev et al, 2011; Hussmann et al., 2011) from tilt of the ground relative to the local gravitational field vector due to the tidal cycle. This sort of measurement has been demonstrated for Earth's tide (e.g., Pillet et al., 1994). These measurements would be sensitive to thermal deformation of the ground, which will occur with almost exactly the same period as the tidal deformation period. Unlike on Earth, where lunar tides and solar effects can be separated after a few cycles, Europa's solar day differs from its sidereal day by only about 1‰. Tidal flexure can also be measured from orbit (e.g., Grasset et al., 2013). Having such a capability on a landed spacecraft could provide improved signal and a longer temporal baseline of measurement.

The frequency and amplitude of planet-scale free oscillations depend on the body's size, the thickness of its ocean-ice layer, and the excitation mechanism. The "football shaped" mode $_0S_2$ was used by both Panning et al (2006) and by Lekic and Manga, (2006) to illustrate possible excitation by tidal and tectonic forcing. Although Europa's radius is a factor of ~4 smaller than Earth's, the frequency of $_0S_2$ is low (~0.1 mHz) compared to the equivalent normal mode on Earth (~0.3 mHz) due to the presence of a water ocean, which mostly controls the $_0S_2$ period. Tectonic excitation of this mode by even a $Mw$=8 ice-quake is likely undetectable (Panning et al., 2006). Tidal forcing might excite $_0S_2$, as investigated for Europa and Enceladus (Lekic and Manga, 2006); this may only be important if the ice shell is thick enough for the frequency of $_0S_2$ to be comparable with some tidal forcing frequencies. Regardless, the frequency of many



normal modes depends on the thickness of the ice-ocean layer. Detecting such normal modes would place a strong constraint on the depth of the ocean and the thickness of the ice crust, although further work is needed to look at excitation of candidate modes other than $_0S_2$.

### 2.2.3. Fluid movements

Fluid transport on Earth generates unique and distinctive seismic sources. Detecting the movement of aqueous fluids within the ice would confirm the presence, at least locally, of hydrological activity and chemical gradients that might be associated with life. If fluids form near the surface—for example, by ridge formation (Dombard et al. 2013), sill emplacement (Michaut and Manga, 2014) or in association with accumulated brines under chaotic terrains (Schmidt et al., 2011)—they can migrate downward into the underlying warm ice and into the ocean (Sotin et al., 2002; Kalousova et al., 2014). Downward transport requires active tidal flexing of the ice to keep the convective adiabat near its melting temperature. Such heating seems likely for tidally heated worlds such as Europa (Tobie et al., 2003; Sotin et al. 2009), but less likely for Callisto and Titan. Alternatively, if tidal dissipation does not sufficiently warm the ice, or if permeability exceeds a threshold required to mobilize fluids and close porosity, fluids may instead move laterally or even upward under hydrostatic pressure toward topographically low regions (Schmidt et al., 2011). Lineaments may generate melt if they move regularly and generate sufficient shear heating (Nimmo et al., 2007b). The slip rate needed to melt ice may be smaller than the $10^{-6}$ m s$^{-1}$ suggested by Nimmo and Gaidos (2002), due to fluids generated at grain boundaries in low-temperature ice (McCarthy and Cooper, 2016).

Near-surface cryovolcanic activity can take two main forms: effusive activity—as suspected at Pluto (Moore et al., 2016), Europa (e.g., Fagents et al., 2000; Kattenhorn and



Prockter, 2014; Quick et al 2013, 2017), and Titan (Tobie et al., 2006)—or eruptive activity as observed at Enceladus (Spencer and Nimmo, 2013) and Europa (Roth et al., 2014; Sparks et al., 2016, 2017). The latter should be accompanied by a characteristic seismic signature that is periodic in nature.

Earth analogues for cryovolcanic activity can be used as a basis for estimating the amplitude and frequency content of associated seismicity. Volcanic degassing episodes (pre-eruptive metastable phases), often last for days to months; similar events on ocean worlds might be expected to produce long-lived seismic signals (Harlow et al., 2004). Another analogue is the seismic activity observed at terrestrial geysers, where a mixture of water, steam, $CO_2$, and sulfur is transported through a conduit system. During the onset of dormant volcanic activity, seismicity is typically composed of micro-quakes (Mw <1) and long-period (LP; >1s) events (Chouet et al., 1994) with quasi-monochromatic signatures. As the eruption progresses, individual LP events merge into a continuous background noise commonly referred to as volcanic tremor. Such events have relatively low amplitudes and so might only be detectable in the vicinity of the seismometer or seismic network. Figure 2 illustrates what seismic information might be detected near an Enceladus plume from volatile transport through a hypothesized conduit system.

## Figure 2 here

In many instances observed in terrestrial volcanoes and geysers, multi-phase flow results in non-linear interactions between the fluid and surrounding solids, giving rise to unique spectral signatures (e.g., Julian, 1994; Kedar et al. 1998). The seismic signal propagated from the source to the seismometer typically generates the appearance of a continuous "hum".



Unlike brittle failures in rock or ice, which have distinct P, S, and surface waves, flow-driven ground motion results from a continuous interaction between the moving fluid and the surrounding rock or ice (Bartholomaus et al., 2015; physics behind this described by Gimbert et al., 2014, 2016). Such ground motions can be caused by a range of fluid states and flow regimes, but commonly appear as continuous low amplitude background vibrations (Fig. 3). A rarely described but especially interesting analogue was observed by Roeoesli et al. (2016), who recorded the seismic signature of water in free fall draining through a glacial moulin from the surface to its base several hundreds of meters below.  The seismic characteristics of such drainage events, which last several hours, bear the hallmarks of flow-driven ground motion displayed in Figure 3. The moulin drainage signal has a sharp onset and abrupt stop, and variable frequency content that closely tracks the measured water level in the moulin.

## Figure 3 here

Trapped liquid layers within the ice, possibly hundreds of meters thick (Schmidt et al., 2011), would have lower sound speeds than surrounding ice, and so would act as a wave guide if sufficiently thick. Such a wave guide was found under the Amery Ice Shelf (H≅500 m) in East Antarctica using cross-correlations of ambient noise between 5-10 Hz.  Incoherent noise makes such measurements sensitive to instrument placement; in the Amery experiment, only a single station showed a clear signal from auto-correlation (Zhan et al., 2014).

### 2.2.4. Oceanic seismic sources

Fluid movements within the ocean might provide distinct seismic signals.  Oceans can also have a large tidal response due to Rossby-Haurwitz waves associated with obliquity and eccentricity (Tyler, 2008, 2014; Chen et al., 2014; Matsuyama, 2014; Kamata et al., 2015;



Beuthe et al., 2016). Such tidally-driven lateral flows may reach speeds comparable to the highest speeds in ocean currents on Earth (>1 m s$^{-1}$; Tyler, 2008, 2014). Resulting dissipation has been parameterized as a term ($Q$) accounting for effects such as boundary-layer friction, form drag, and transfer of momentum to the overlying ice. Seafloor topography causing such dissipation could include mountains with heights exceeding Earth's seamount heights (8 km; Wessel et al., 2010) owing to lower gravity, and by analogy to mountain features on Io (Bland and McKinnon, 2016). The ice shelf's tidal displacement ($u_r$~30 m for Europa; Moore, 2000) and thickness variation with latitude (Nimmo et al., 2007a) also constitute sources of topographic variation with time.

The main source of seismic noise on Earth is the ~3-10 s background noise caused by opposing travelling ocean waves with overlapping frequency content. This noise source, known as ocean microseism, results from a second-order interaction between surface gravity waves. Frequency doubling means the seismic signals have half the period of the originating swell (Longuet-Higgins, 1950). Ocean microseism can be enhanced if resonance occurs within the water column, as theorized by Longuet-Higgins (1950) and demonstrated by Kedar et al (2008, 2011; see also Gualtieri et al., 2013; Ardhuin et al., 2015). This swell has a typical period of 13 s, resulting in seismic signal excitation around 6 s. For a free surface ocean, the first mode of acoustic resonance takes place at ¼ wavelength. Given a speed of sound in water of 1500 m s$^{-1}$, the first mode of a ~6 s wave would take place when the ocean is 2.25 km deep. The pressure waves resonantly loading the ocean floor generate Scholte waves, surface waves that propagate along the water-rock interface and can be observed thousands of kilometers away from the source (Stutzmann et al., 2009).



Ocean worlds capped by ice lack the surface winds that are the source of ocean microseisms on Earth, but vertical motions driven by global tidal flexure may substitute as a source. The first mode of resonant excitation would be ½ the acoustic wavelength, and the expected resonant period, $T$, would be:

$$T = 2h/c$$

where $h$ is the ocean depth and $c$ is the speed of sound in water. Assuming $c$=1500 m/s, the resonant period could vary from 40 s for a 30 km ocean to 227 s for a 170 km ocean. If ocean microseism occurs in ocean worlds, Scholte surface waves along the water-ice interface might be detected by a very broad-band seismometer at the ice surface.

Movement of liquid in Titan's major sea Kraken Mare due to tides, winds and other effects is likely to lead to 10s of cm of local level change (Lorenz & Hayes, 2012), creating pressure variations on the sea-floor of around 100 Pa. This is a similar pressure fluctuation as is measured on the deep sea-floor of the Earth (e.g. Cox et al. 1984), which excites the significant secondary microseism there (Longuet-Higgins 1950, Kedar et al. 2008). Thus, ocean-generated microseisms might also be detected on Titan by a seismometer close to the sea shore,

Another possible oceanic source is the low-level excitation of normal modes by motion in the ocean. On Earth, seismic normal modes at frequencies ~10 mHz are excited at a nearly constant level by interaction of the ocean with the continental shelf (Webb, 2007) and comparable excitations have been proposed for Mars and Venus (Kobayashi and Nishida, 1998). Turbulent oceanic flows in Europa (Soderlund et al., 2014) plausibly produce acoustic transmissions through the ice through dynamic pressure variations at the base of the ice shell that are comparable to those from the estimated global background noise due to fracturing at



frequencies from ~10 to 100mHz (Panning et al., 2017); modeling how well this mechanism excites specific modes will depend strongly on the frequency and wavelengths of the turbulent motion. Although the excitation mechanism for an ocean entirely covered in ice will undoubtedly differ and will require future study, the possibility of a constant excitation of normal modes is intriguing as it may shed light on both internal structure constraining the composition of the ocean (Vance et al. 2017) and oceanic processes governing the degree of material and heat exchange (Zhu et al. 2017).

### 2.2.5. Seismic sources in the rocky interior

In the rocky interior, more radiogenic heat or residual heat of formation should promote viscoelastic deformation, leading to stronger tidal heating (e.g. Showman and Malhotra, 1997; Cammarano et al., 2006; Bland et al., 2009). Both sources of heat were much stronger earlier in the solar system's history. The remnant heat is not quantified, creating uncertainty in the contemporary level of seismic activity in the rocky interiors. The thermal state and density of the rocky mantle thus indicate the nature of any continuing volcanic activity (Barr et al., 2001), water rock interaction (e.g., due to serpentinization and radiolysis), and the corresponding flux of reducing materials ($H_2$, $CO_2$, $H_2S$ $CH_4$; McCollom, 1999; Hand et al. 2009; Holm et al., 2015; Travis and Schubert, 2015; Vance et al., 2016; Bouquet et al., 2017).

Events in the rocky interior (and to a lesser extent events in the ice layer) will excite Scholte-waves at the rock-ocean boundary (Stähler et al. 2017). Their dispersion is used in exploration seismology and ocean acoustics to estimate the shear velocity of the sea floor sediment (Jensen & Schmidt, 1986). The wave coda—diffuse energy arriving after a clearly defined seismic phase—constrains the amount of scattering, especially in the receiver-side



crust, and thereby the heterogeneity within the ice layer. On the Moon, this has been used to derive the existence of the megaregolith layer (Goins et al., 1981; Blanchette-Guertin et al., 2012). In the ice shells of ocean worlds, it would constrain the distribution and extent of trapped fluid layers.

Hydrothermal activity could generate seismoacoustic signals that travel through the internal ocean and ice where they could be intercepted by a surface seismometer. For this and all oceanic measurements, the impedance mismatch between ocean and ice would limit the practicality of such measurements except for very high magnitude sources (Panning et al., 2006).

## 3. Mission Requirements and Constraints

Seismic investigations will need to focus on how to most effectively enhance the mission's astrobiology objectives while levying minimal requirements on limited mass, power and bandwidth resources.   The seismometer's sensitivity, dynamic range, and bandwidth depend on the size of the proof mass, the damping of the suspension system, and the noise of the readout electronics (Lognonné and Pike, 2016).  Consequently, sensitivity adds complexity, mass, and cost. The challenge for planetary seismic exploration, and for ocean worlds in particular, is to devise seismometer systems that deliver science within tight mission constraints.

To date the only extra-terrestrial seismic studies have been conducted on the Moon, by the *Apollo* missions (Goins et al., 1981), on Mars by the *Viking* landers (Anderson et al., 1977; Goins and Lazarewicz, 1979), and on Venus by the *Venera* 13 and 14 landers (Ksanfomiliti et al., 1982). Lognonné & Johnson (2007) and Knapmeyer (2009) provide thorough reviews of results



from these experiments. The *Apollo* missions established a sensitive seismic network for a global study of the deep lunar interior, and also conducted a shallow active study using small explosives and geophones to study the shallow (~1km) lunar subsurface.  The seismic network set up by the Apollo astronauts in multiple lunar landings was sensitive though narrow band (0.2-3 Hz) and detected over 7 years about ten thousand tidally triggered quakes that are presumed to occur at the base of the lunar lithosphere (~800-1000 km depth; Frohlich and Nakamura, 2009, Weber et al., 2009), as well as about 1000 meteorite impacts (Lammlein et al., 1974; Lognonné et al., 2009).  These data constrain the depths of the lunar crust (e.g. Vinnik et al., 2001, Khan and Mosegaard, 2002; Chenet et al., 2006), its mantle (Khan et al, 2000; Lognonné et al., 2003; Gagnepain et al.2006), and its core (Weber et al., 2011, Garcia et al., 2011).

Both *Viking* landers carried seismometers to Mars, but only the one on the *Viking* 2 deck operated successfully. The *Viking* data are often dismissed as having limited value due to the susceptibility of the lander mounted instrument to wind noise, but they helped constrain global seismicity on Mars (Anderson, 1976; Anderson et al.*, 19*77).  Similarly, the Soviet *Venera* 13 and 14 landers were hampered by wind and spacecraft noise, but, they set upper bounds on the magnitude of microseisms, and potentially observed two microseism events with amplitudes of 0.8 µm or less, within 3000 km of *Venera* 14 (Ksanfomaliti et al., 1982).

*InSight* (Interior Exploration using Seismic Investigations, Geodesy and Heat Transport), a Discovery-class mission, will place a single geophysical lander on Mars to study its deep interior in November of 2018 (Banerdt et al., 2013). *InSight*'s goals are to (1) understand the formation and evolution of terrestrial planets through investigation of the interior structure and



processes of Mars; and (2) determine the present level of tectonic activity and impact flux on Mars. To avoid the coupling problems encountered by Viking, *InSight* will place a top-of-the-line broad-band high-dynamic-range seismometer (Lognonné and Pike, 2015) on the martian surface to ensure optimal coupling, and will effectively build a seismic vault around it by deploying a wind and thermal shield around the seismometer.

A studied future mission to the Moon, the *Lunar Geophysical Network* (LGN) is identified as a high-priority New Frontiers class mission in the Planetary Science Decadal survey, and seeks to understand the nature and evolution of the lunar interior from the crust to the core. The *LGN* requires a very broad band seismometer, ten times more sensitive than the current state of the art (Shearer and Tahu, 2010; Morse et al., 2010).

*InSight* and the *LGN* represent the high sensitivity end-member on a spectrum of planetary seismic exploration concepts. The requirement to operate with similar sensitivity to instruments on Earth is underscored by the fact that both the Moon and Mars are less active than Earth. As noted in studies of potential landed Europa missions (Europa Science Definition Team, 2012; Pappalardo et al. 2013; Hand et al. 2017), these requirements can be substantially relaxed in the study of ocean worlds, which are expected to be more seismically active. This means the seismometers delivered to the surfaces of the solar system's ocean worlds can be smaller, lighter, and less complex than *InSight*'s broad-band seismometer.

### 3.1. Environmental Limitations on Lifetime

Planning for missions to different moons dictates a diverse design space covering duration, mass and power allocation, cumulative radiation tolerance, and deployment capabilities. Table II shows the range of surface temperatures and estimated surface irradiation



for current candidate ocean worlds (Nimmo and Pappalardo, 2016). Radiation is deemed to be mission limiting on Europa (Pappalardo et al., 2013), but is less problematic elsewhere, possibly enabling longer-duration investigations.

## Table II here

### 3.2. Dynamic Range and Sensitivity

Potential seismic sources span a broad range of signal strengths and frequencies (Fig. 1), which calls for measuring ground motions from nanometers to the tidal bulge of tens of meters (ten orders of magnitude). The natural noise environment on a given world can be estimated by constructing a catalogue of events for realistic simulations of likely sources, and assuming a logarithmic relation between of magnitude and occurrence (Gutenberg and Richter 1944):

$$logN(M_W) = a - bM_W$$

Analyses of this type performed for fractures at Europa's surface (Panning et al 2017) suggest a low noise floor is needed, with instrument performance exceeding that of high-frequency geophone instruments.

Scattering in the porous surface regolith of an airless body may be significant, and will further increase the needed sensitivity of any potential instrument. Pore size is a proxy for scattering, for ground penetrating radar as well as for seismology. The depth and porosity of Europa's warm regolith are probably small (Aglyamov et al., 2017), and so scattering may be less consequential. In the older bodies with less tidal heating, especially Callisto, the regolith may be on the order of 10 km thick (McKinnon 2006), thus necessitating a greater sensitivity



and longer baseline of measurement similar to those being considered for the Lunar Geophysical Network (Shearer and Tahu, 2010).

### 3.3. Lander Noise

As reviewed by Lorenz (2012), landers can generate spurious signals. The seismometer on *Viking* was mounted on the lander deck, and wind noise rocking the lander on its legs was the dominant noise source. On Europa and other ocean worlds lacking atmospheres, this should not be a concern. Thermal creaking of metal structures, and movement of residual propellant in fuel tanks in response to changing temperatures, caused detectable noise on *Apollo*, even though those seismometers were placed on the lunar ground some distance away. The *Apollo* 11 instruments, 16 m from the lander, saw much more of this noise (e.g., Latham et al., 1970, 1972), and the instrument package on subsequent missions was deployed more than 100m from the landers.

Solar flux decreases with distance from the Sun (at Jupiter 25x and at Saturn 80x less than at Earth). Thermal gradients should be less severe than at the Moon, and thus lander-induced noise should be weaker on a per-unit-mass basis. Beyond this expectation, however, lander noise is difficult to predict in advance. Some mechanical operations, such as antenna articulation, regolith sample acquisition by a robot arm or similar device, and sample analysis operations such as grinding or valve actuations, may generate signals detectable by an on-board or even nearby seismometer. Indeed, the seismometer data may be useful in diagnosing any off-nominal behavior of such equipment. However, it would be desirable to define extended 'quiet' periods when such operations are vetoed to allow seismic observations with the minimum lander background, and that such periods be distributed around the diurnal cycle



to allow for characterization of the ambient seismic activity as a function of the tidal cycle. As discussed above, installing a seismometer should be much simpler on airless bodies, and so such features may not be needed.

### 3.4. Studied Mission Implementations and Candidate Instrumentation

Lander concepts for Europa were studied in 2012 (Europa Science Definition Team, 2012; Pappalardo et al., 2013) and 2016 (Hand et al. 2017). The first of these would address habitability using a set of six broadband 3-axis seismometers (0.1-250 Hz) based on the Exomars and Insight instruments (Pappalardo et al., 2013). Detailed objectives were to "understand the habitability of Europa's ocean through composition and chemistry," "characterize the local thickness, heterogeneity, and dynamics of any ice and water layers," and "characterize a locality of high scientific interest to understand the formation and evolution of the surface at local scales." Lander lifetime was limited by the combination of Europa's surface radiation (Table II) and constraints on shielding mass to fewer than 30 days spent on the surface. The average unshielded dose of 6-7 rad $s^{-1}$ varies with latitude and longitude on Europa's surface, as the incident flux of high-energy electrons (0.01-25 MeV) from Jupiter's ionosphere centered around Europa's trailing hemisphere has a longitudinal reach around Europa that diminishes with increasing energy (Paranicas et al., 2009; Patterson et al., 2012). Thus, radiation may be less limiting for a lander placed on the leading hemisphere. A Ganymede lander would also need to contend with intense radiation, albeit 20x less than that at Europa. The trailing hemisphere may be more irradiated, similar to Europa, but Ganymede's intrinsic magnetic field may protect from irradiation around the polar regions.



Titan has the advantage that its thick atmosphere makes it easy to deliver instrumentation to the surface by parachute: the 2007 NASA Titan Explorer Flagship study included a Pathfinder-like lander equipped with a seismometer (Leary et al, 2007).   The thick atmosphere also minimizes diurnal temperature changes that can generate local disturbances.  On the other hand, meteorological pressure variations acting on the surface, and local wind stresses, may generate both ground deformations (*e.g.*, Lorenz, 2012) that increase the need for strong coupling to the ground and might be the source of atmospheric excitation of surface waves and normal modes through solid planet-atmosphere coupling (Kobayashi and Nishida, 1998; Lognonné et al., 2016).

Seismology has also been included as part of studied penetrator missions to Europa (e.g., Gowen et al., 2011, Jones et al., 2016), Ganymede (Vijendran et al., 2010), Callisto (Franqui et al., 2016), and Enceladus and Titan (Coustenis et al., 2009), with the goal of confirming and characterizing the ocean, probing the deeper interior structure, and assessing the level of seismic activity. A seismometer under development at the time (Pike et al., 2009) showed promise for withstanding the required 5-50 kgee impact deceleration.

As recently demonstrated by Panning et al. (2017), the measured noise floor of the microseismometer that was successfully delivered for the InSight Mars 2018 mission demonstrates a sufficient sensitivity to detect a broad range of Europa's expected seismic activity. While this seismometer is designated as "short period" (in comparison to the CNES-designed very broadband (VBB) seismometer), the SP provides a sensitivity and dynamic range comparable to significantly more massive broadband terrestrial instruments. The sensor is micromachined from single-crystal silicon by through-wafer deep reactive-ion etching to



produce a non-magnetic suspension and proof mass with a resonance of 6 Hz (Pike et al., 2014). The SP is well suited for accommodation on a potential Europa Lander. It is robust to high shock (> 1000 g) and vibration (> 30 grms). The sensor has been tested as functional down to 77K, below the lowest expected temperatures on Europa's surface. All three axes deliver full performance over a tilt range of ±15° on Mars, allowing operation on Europa without leveling. The SP operates with feedback to automatically initiate power-on of the electronics, achieving a noise floor below 1 ng/√Hz in less than a minute. The total mass for the three-axis SP delivery is 635 g while the power requirement is 360 mW.

### 3.5. Example Science Traceability Matrix

Table III shows a candidate traceability matrix for Europa, developed from the detailed traceability matrix of *Insight* (Banerdt, *personal communication*) with reference to the features reviewed here and summarized in Fig. 1. The Science Objectives (column 2) meet the overall Science Goal to "Determine Europa's habitability, including the context for any signatures of extant life." Observable features, the associate physical parameters, and derived properties are contained in the Measurement Requirements (columns 3 and 4), which link to estimated Instrument Performance Requirements (column 5).

Similar goals and associated objectives, measurements, and performance requirements can be developed for other ocean worlds. On other icy ocean worlds, the level of activity may be only slightly less, or perhaps more in the case of Enceladus or Titan, so the anticipated sensitivity and dynamic range would be similar. However, requirements will diverge based on the presence or absence of different phases of high pressure ice (Stähler et al., 2017; Vance et al. 2017), and the differing extent and nature of present day activity (Panning et al., 2017). As



noted by Vance et al. (2017), Ganymede is the only world likely to possess substantial amounts of ice VI. Titan may lack high pressure ices, and should include investigations of the atmosphere and lakes. Callisto probably has the lowest level of seismic activity and strongest scattering in its regolith, and so would require a longer-lived and more sensitive investigation similar to the of the *Lunar Geophysical Network*.

## Table III here

## 4. Conclusions

Seismology is the best tool for remotely investigating possible "vital signs", ground motions due to active fluid flow, in ocean worlds, yet only a handful of possible seismic sources have been considered to date. Detecting fluid-related seismic signatures similar to those on Earth would provide additional key information for constraining transport rates through the ice, and associated redox fluxes, and locating possible liquid reservoirs that may serve as habitats. Seismic activity in tidally forced icy ocean worlds is likely to exceed that recorded at Earth's Moon and expected at Mars. We document design challenges for potential future missions, with reference to prior mission implementations and studies, and to recent studies of signal strength and propagation.

## 5. Acknowledgements


We thank for their helpful input Sridhar Anandakrishnan, Bruce Banerdt, Jason Goodman, and Jennifer Jackson. This work was partially supported by strategic research and technology funds from the Jet Propulsion Laboratory, Caltech, and by the Icy Worlds node of NASA's Astrobiology Institute (13-13NAI7_2-0024). The research was carried out at the Jet






## 6. Author Disclosure Statement.

No competing financial interests exist.

| | Mean Radius in km | Bulk Density in kg m$^{-3}$ | Moment of Inertia (C/MR$^2$) | Inferred H$_2$O thickness in km | Rotation Period In hrs | Radiogenic Heat in GW | Tidal Dissipation in GW | Seismic Energy Release in GW (%) |
|---|---|---|---|---|---|---|---|---|
| **Earth**[a] | 6371 | 5514 | 0.3307 | 3.5 | 24 | 31,000 | 2636±16 | 22 (0.8) |
| **Moon**[a,b] | 1738 | 3340 | 0.3929 ± 0.0009 | n/a | 672 | 420 | 1.36±0.19 | 6x10$^{-7}$ (0.4x10$^{-5}$) |
| **Mars**[c] | 3397 | 3933 | 0.3662 ± 0.0017 | n/a | 24.6230 | 3,300 | -- | ? |
| **Europa**[d,e,f] | 1565.0± 8.0 | 2989±46 | 0.346 ± 0.005 | 80-170 | 84.4 | 200 | <10,000 >1000 (ocean) 1-10 (rock) | ? |
| **Ganymede**[e,f,g] | 2631.17 ±1.7 | 1942.0±4.8 | 0.3115±0 .0028 | 750-900 | 171.7 | 500 | <50 >1 (ocean) | ? |
| **Callisto**[e,f] | 2410.3± 1.5 | 1834.4±3.4 | 0.3549±0 .0042 | 350-450 | 400.5 | 400 | <20 >4 (ocean) | ? |
| **Titan**[e,h,i] | 2574.73 ±0.09 | 1879.8±0.2 | 0.3438±0 .0005 | 500-700 | 382.7 | 400 | <400 >11 (ocean) | ? |
| **Enceladus**[e,h,j] | 252.1±0. 1 | 1609±5 | 0.335 | 60-80 | 32.9 | 0.3 | <20 >10 (ocean) | ? |

**Table I.** Mean radii in km, bulk density in kg m$^{-3}$, gravitational moment of inertia (C/MR$^2$),

corresponding range of possible thicknesses of the upper H$_2$O-rich part of the body (km),

rotation periods (in hrs), radiogenic and tidal heat in GW, and seismic energy release (where

known) in GW and as a percent of tidal dissipation. Moment of inertia values (and

corresponding H$_2$O thicknesses) may be over (under) estimated by up to 10% in Callisto (Gao

and Stevenson 2013). Estimates of tidal heating in known ocean worlds[e], especially in the

ocean, are highly uncertain, varying with obliquity; ocean depth, viscosity, and interface

friction; and dissipation of the companion body. Values are given to one significant figure. In ice

they are upper limits for a fluid body (k$_2$=3/2), and are roughly ten times less accounting for

likely values of rigidity.

a) Williams et al., 2001 b) Goins et al., 1981; Williams et al., 1996; Siegler and Smekrar, 2014 c)

Folkner et al., 1997, Nimmo and Faul, 2013 d) Tobie et al. 2003, Hussmann et al., 2006; Vance



et al., 2007 e) Chen et al., 2014; Tyler, 2014 f) Schubert et al. 2004 g) Bland et al., 2015 h) Jacobson et al., 2006 i) Iess et al., 2012; Gao and Stevenson, 2013 j) Howett et al., 2010; Spencer and Nimmo, 2013; Iess et al., 2014, McKinnon, 2015; Beuthe et al., 2016.



| Candidate Ocean World | Surface Temperature (K) | Surface Ionizing Radiation (mW m$^{-2}$) |
|---|---|---|
| Europa[a,b] | 50-132 | 125 |
| Ganymede[b,c] | 70-152 | 6 |
| Callisto[b,c] | 80-165 | 0.6 |
| Enceladus[d,e] | 33-145 | <6 |
| Titan[e,f] | 90-94 | <6 |

**Table II.** Temperature ranges and radiation fluxes likely to be encountered by a seismometer at the surface of an ocean world.

a) Spencer et al.,1999; Rathbun et al, 2010; Prockter and Pappalardo, 2014 b) Johnson et al., 2004 c) Orton et al., 1996 d) Howett et al., 2011 e) Cooper et al., 2009 f) Mitri et al., 2007; Jennings et al. 2016.



| Science Goal | Science Objectives | Scientific Measurement Requirements | | Instrument Performance Requirements [Hz] / [m/s²/√Hz] |
| --- | --- | --- | --- | --- |
| | | Observables | Physical Parameters & *Derived Properties* | |
| Determine Europa's habitability, including the context for any signatures of extant life | Characterize the [global/local] thickness, heterogeneity, and dynamics of the ice | Surface wave dispersion curves (for thickness) and body wave coda (for scattering/heterogeneities) | Ice thickness, sound speed, attenuation, patterns of activity (*temperature structure, impurities, locations of energy release*) | 0.05-10Hz $10^{-8}$ m/s²/√Hz |
| | Characterize the [global/local] thickness, heterogeneity, and dynamics of fluid layers within the ice | Seismic body and surface wave arrival times | Ice thickness, sound speed, attenuation (*temperature structure, impurities*) | 0.05-10Hz $10^{-8}$ m/s²/√Hz |
| | | Fluid induced tremor | Fluid flow rates and volumes (*potentially habitable regions*) | 0.1-10Hz $10^{-8}$ m/s²/√Hz |
| | Characterize the [global/local] thickness, heterogeneity, and dynamics of the ocean | Body wave arrival times | Ocean depth, structure, sound speed (*salinity, structure, temperature*) | 0.1-10Hz $10^{-7}$ m/s²/√Hz |
| | Determine the composition and structure of Europa's rocky mantle and the size of any metallic core | Body wave arrival times | Mantle and core depth, sound speed (*mineralogy, structure, temperature*) | 0.1-10Hz $10^{-8}$ m/s²/√Hz |
| | | Tide induced displacement | Radial mass distribution and presence of melt in the core | $10^{-5}$Hz $10^{-4}$ m/s²/√Hz |
| | | Free Oscillations | Radial mass distribution and presence of melt in the core | <$10^{-2}$Hz $10^{-9}$ m/s²/√Hz |

**Table III. Candidate traceability matrix for a seismic investigation of Europa.**



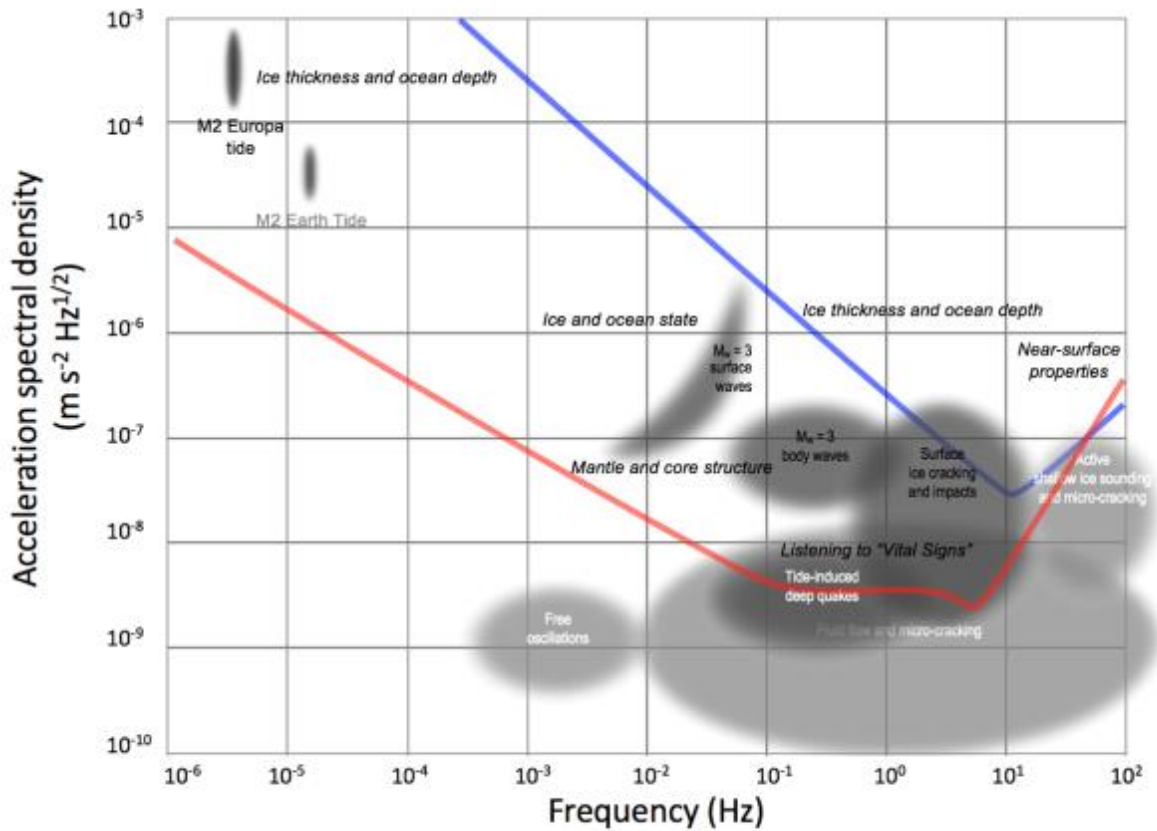

**Figure 1:** Europa is expected to be seismically active (Panning et al., 2006; Lee et al., 2003). A sensitive, broad-band, high-dynamic-range seismometer (red, Pike et al. 2016) could detect faint seismic signals associated with ice-quakes, and fluids flow within and beneath the ice crust to constrain chemical, and thermal structures and processes. The performance of a 10 Hz geophone is shown for comparison.



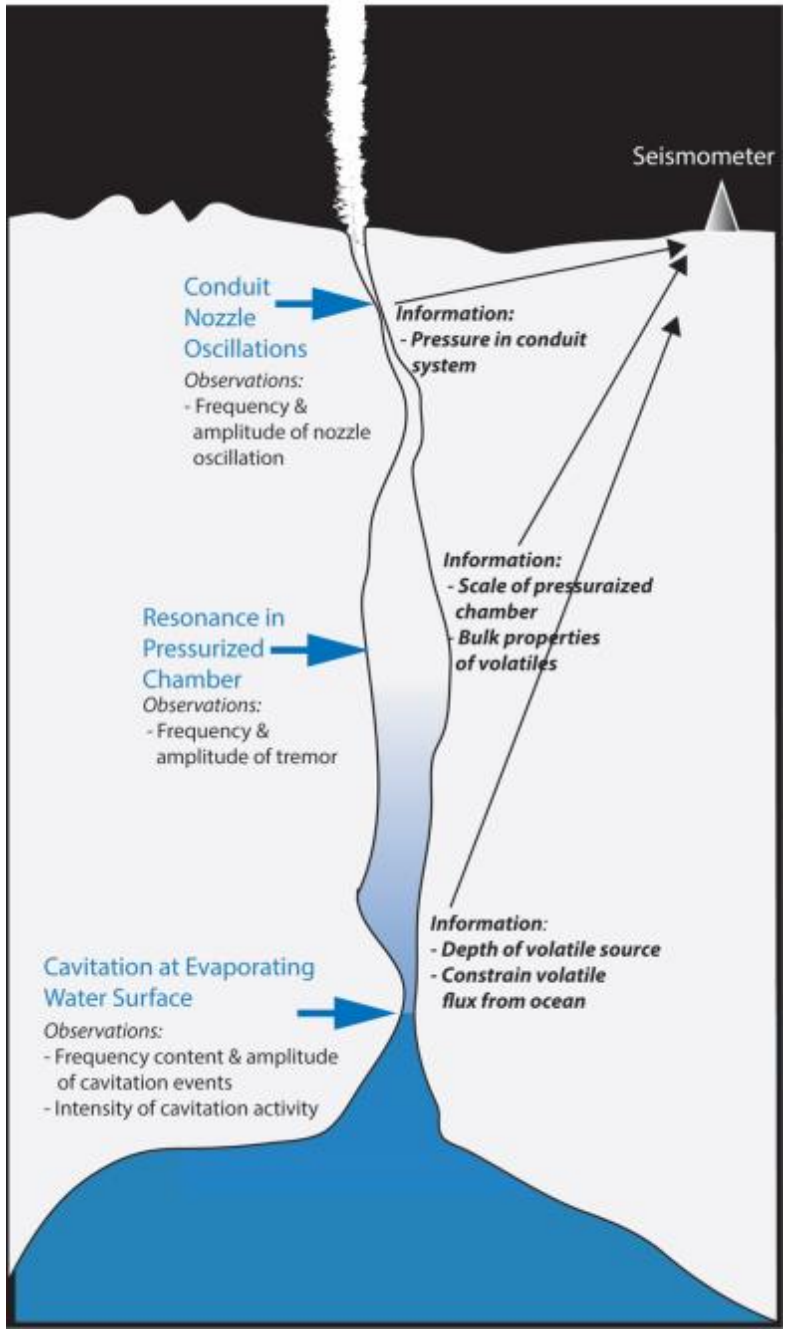

**Figure 2.** Potential seismic sources (blue text) how they might be used to constrain the physical properties and parameters of a cryovolcano system (after Spencer and Nimmo, 2013). The hypothesized seismic sources of cavitation, chamber resonance, and nozzle oscillations are based on terrestrial analogues of geysers (Kedar et al., 1996; 1998), volcanoes (Chouet et al., 1994) and volcanic nozzles (Kieffer, 1989) respectively. The geometry depicted here resembles that



suggested for Enceladus by Schmidt et al. (2008), but a simpler "slot" geometry may occur instead (Kite and Rubin, 2016).

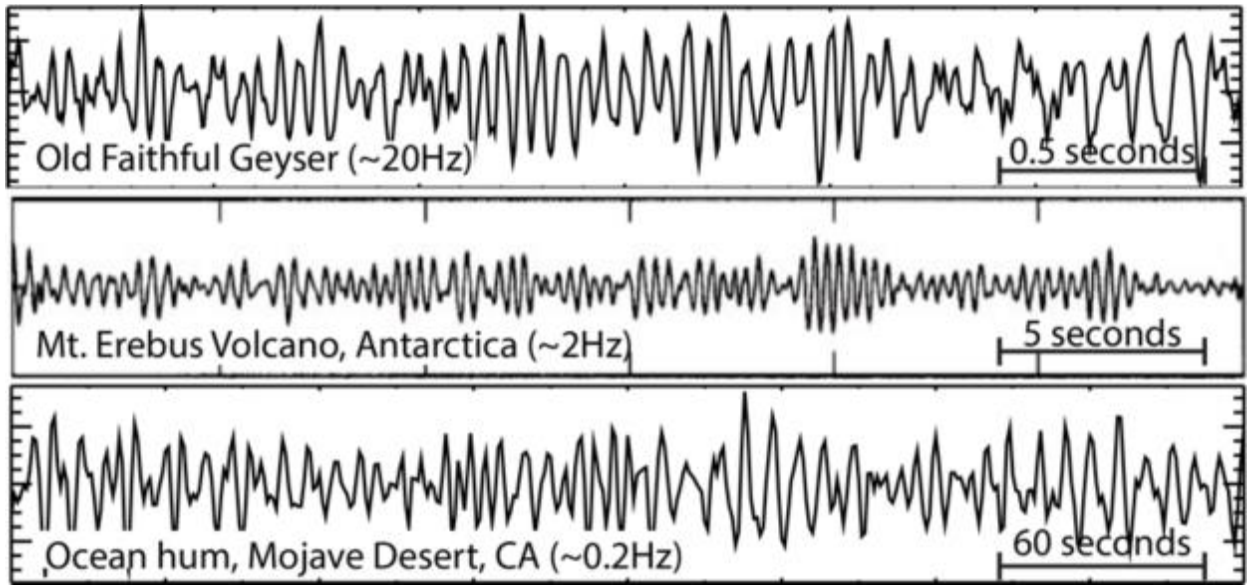

**Figure 3:** Fluid flow induced ground motion is recognizable by its appearance as continuous background oscillation. Different sources have distinct frequencies and variations in amplitude, as shown in these example seismograms. **Top**: Geyser ground motion (~20 Hz) generated by steam bubble collapse at Old Faithful (Kedar et al. 1996, 1998); **Middle**: Volcanic tremor (~2 Hz) generated by magma and hot gas rising in the Mt. Erubus volcano's conduit system (Rowe et al. 2000); **Bottom**: Acoustic hum (~0.2 Hz) generated by wave-wave interaction is detectable globally (Longuet-Higgins 1950, Kedar et al., 2008).